\documentclass[twocolumn]{article}

\author{Collin M. Cunningham}

\usepackage{graphicx, subcaption, amsmath, amsfonts}
\usepackage[top=2cm, bottom=4cm, left=2cm, right=2cm]{geometry}
\usepackage{fancyhdr}

\makeatletter
\renewcommand{\maketitle}{\bgroup\setlength{\parindent}{0pt}
\begin{flushleft}
  \textbf{\@title}

  \@author
\end{flushleft}\egroup
}
\makeatother

\title{\Large Trigonometric Extension of the Geometric Correction Factor: Prototype for adding precision to adaptive ray tracing in ENZO}
\date{}
\author{%
Collin M. Cunningham$^{1}$: Under the supervision of John H. Wise, Ph.D.$^{2}$\\
$^{1}$Center for Computational Cosmology - School of Physics - Georgia Institute of Technology, Atlanta Georgia, USA\\
    $^{2}$Center for Relativistic Astrophysics - School of Physics - Georgia Institute of Technology, Atlanta Georgia, USA\\
    \underline{$^{1}$ccunningham9@gatech.edu}\\
     \underline{$^{2}$jwise@physics.gatech.edu}
}

\begin{document}
\twocolumn[
  \begin{@twocolumnfalse}
	\maketitle
        \vspace{.5cm}
    \begin{abstract}
    \vspace{.2cm}
    
      In this paper, we describe a method designed to add precision to radiation simulations in the adaptive mesh refinement cosmological hydrodynamics code ENZO. We build upon the geometric correction factor described in \textit{ENZO+MORAY: radiation hydrodynamics adaptive mesh refinement simulations with adaptive ray tracing} (Wise and Abel 2011) which accounts for partial coverage of a ray's solid angle with a cube. Because of this geometric mismatch in the methods to approximate this, there are artifacts in the radiation field. Here, we address the two-dimensional extension, which acts as a sufficient estimate of the three-dimensional case and, in practice, the Hierarchical Equal Area isoLatitude Pixelization of the sphere (HEALPix) (Gorski 2005). We will demonstrate the value of an extension to the geometric correction factor and lay the groundwork for a future implementation to ENZO to improve simulations of radiation from point sources. 
      \vspace{1.5cm}
    \end{abstract}
  \end{@twocolumnfalse}
]

\section{Introduction}
In ENZO, rays are generated from point sources of radiation, which then travel radially outward. The ray normals are calculated from HEALPix (Gorski 2005). Some number of photons are associated with each ray. Photons are subsequently deposited into grid cells based on the geometric overlap between the volume and the computational grid. These absorbed photons ionize and heat gas. Because of the geometric mismatch between the volumes associated with the rays and a Cartesian grid, there are artifacts in the radiation field. We are able to remedy many of these instances by adding precision to the radiation field in a rather simple way: a Trigonometric extension of the geometric Correction Factor (herein TegCF). A prototype for TegCF is the foundation of this paper, which will serve as an indication of the value of implementing a full version.

We begin by simplifying this problem in several ways. Firstly, we will only be considering two-dimensional space, mapping rays onto the Cartesian plane. Therefore, instead of each ray having a volume, each ray will have an area. This allows us to use a series of trigonometric calculations to deduce the proportion of photons to deposit, wherein lies the mechanics of TegCF. Also, we assume we are far from the point-source of radiation; thus, rays will be parallel. This reduction in complexity suffices as a reasonable approximation of the in precision added by TegCF, because the current method in ENZO (Wise Abel 2011) operates under these assumptions as well on a given cell in two dimensions. Therefore, we can directly compare examples of the two methods in small discrete cases. 

Currently, only the cells with which the ray directly intersects can accept radiation. By depositing radiation in neighboring cells, we can add precision in these calculations and retroactively get a better estimate of how much radiation to deposit in the intersectional cell. The accuracy added to the intersectional cell arises from the fact that the current method uses an approximation of area and does not compute it directly. This will, in turn, fix some of the artifacts in computations and make radiation simulations generally more accurate.

\section{The TegCF method}

In this section, we outline the explicit methodology behind TegCF. We limit our discussion to the intersection of the ray with one cell. This rectangle area is called a \textit{pixel} (figure 1). Also, we assume the ray enters on the left side. As is customary in Cartesian coordinates, let each cell have a width and height of $1$ (for implementation, this will be replaced by some $\Delta c$). We name the height of the point where the ray enters the cell $Q$, such that $Q \in [0,1)$. Each ray will also have an angle $\phi$, where $\phi$ is the angle between the ray and the $x$-axis, $\phi \in [0, \pi/2)$. Let the width of the ray be $L_{pix}$ upon entering the cell and remain constant until the ray exits the cell. We limit width by $L_{pix} \le 1$ ($= \Delta c/2$ for implementation). This clearly does not address many possible ways a ray can enter a cell. The extension from this to all possible entry cases is given at the end of this section.

We want to turn our attention to the covering fraction of neighboring cells, labeled 1 through 9 in figure 1. Let $f_i$ denote the covering fraction of cell $i$. Given the assumptions above, note that $f_3 = f_7 = 0$. This follows from the restrictions on $\phi$.

\begin{figure} [!h]
\centering
\includegraphics[totalheight=0.3\textheight]{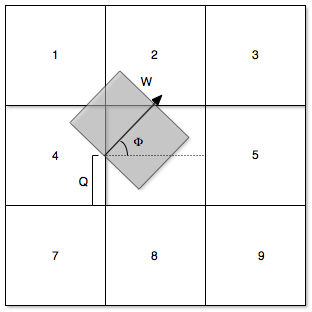}
\caption[short]{Illustration of our simplification to single cell, single ray. The grey shaded area is called a \textit{pixel}.}
\end{figure}

Covering fractions in general will be found using the following formula:
$$
f_i = \frac {A_i} {\mathrm dr \times L_{pix} }\ ,
$$
where $\mathrm d r$ is the distance travelled by the ray-vector from its entry point in cell 5 to its exit point on the boundary of either cell 2 or 6, and $A_i$ is the area of the overlap of the pixel and the cell. 

\subsection{Cell covering areas}
We will briefly address how we found the $A_i : i = 1,2,4,6,8,9$. The accompanying source code for the prototype gives further insight. The following aggregate variables are given for brevity.
\begin{table}[htbp]
\centering
 \begin{tabular}{c c} 
 \multicolumn{2}{c}{Commonly used variables} \\ [0.5ex] 
 \hline
 $W$ & $L_{pix}/2$ \\ 
 $Y$ & $W \tan \phi$  \\
 $Z$ & $W\sec \phi$\\
 $A$ & $W\cos \phi$\\ 
 $U$ & $W\sin \phi$\\ 
$L$ & $\mathrm dr$\\  [1ex] 
 \hline
 \end{tabular}
\end{table}
Some of these variables are not used in the formulas given here, but we give them anyway as an aide for parsing the protoype's code.

\subsubsection{Cell 1} $Z + Q$ must be greater than 1, else the area is zero.
\begin{itemize}
\item  $Q +A  \le 1$: We have a triangle with sides $Q + Z - 1$ and $(Q + Z - 1)\cot \phi$. The area is $$A_1 = (Q + Z - 1)^2 )\cot \phi/2$$
\item $Q + A > 1$: We find this area by subtracting out the extension of the upper side of the ray from the previous iteration.  The area is $$A_1 = \frac 12[(Q + Z - 1)^2 )\cot \phi- (Q + A -1)^2(\cot \phi + \tan \phi)]$$
\end{itemize}

\subsubsection{Cell 2} There are two primary scenarios here, the one in which the ray exits through cell 2 and when it exits through cell 6. Each of these cases has two sub cases, one for the case of a triangle and the other for a quadrilateral.
\begin{itemize}
\item $\tan \phi + Q \ge 1$: $Z + Q \le 1$: $$A_2 = W^2\cot \phi/2$$
\item $\tan \phi + Q \ge 1$: $Z + Q > 1$: $$A_2 = W^2\cot \phi/2 - (Q + Z -1)^2 \cot \phi /2$$
\item $\tan \phi + Q < 1$: $Z + Q \le 1$: Let $h = (1 - U)\tan \phi + Q + Z -1$. 
$$A_2 = h^2 (\cot \phi + \tan \phi)/2
$$
\item $\tan \phi + Q < 1$: $Z + Q > 1$:
$$A_2 = h^2 (\cot \phi + \tan \phi)/2 - (Q + Z -1)^2 \cot \phi /2
$$
\end{itemize}

\subsubsection{Cell 4}
\begin{itemize}
\item $Q + A \ge 1$: The covering area of Cell 4 is a triangle with height $(1-Q)$ and base length $(1-Q) \cot \phi$. Therefore, we have a covering area of $$A_4= (1-Q)^2 \cot \phi/2$$

 \item  $Q+ A < 1$: $Z+Q \le1$: We again have a triangle but with sides $W$ and $Y$ implying $$A_4=WY/2$$
 
\item $Q+ A < 1$: $Z+Q >1$: The remaining case results in a quadrilateral. We can deduce that $B = W \sin \phi$, $C = 1 - A - Q$, $E = C \cot \phi$. This results in $$A_4 = (AB+CE)/2 + C(B-E)$$
\end{itemize}

\subsubsection{Cell 6} These cases correspond to where the ray exits as in cell 2. There is a simplification by subtracting out cell 9 in the case where the ray exits through cell 6.
\begin{itemize}
\item $\tan \phi + Q \le 1$: 
$$A_6 = YW/2 - A_9
$$
\item $\mathrm dr _x + U > 1$ (checks whether there is any overlap in the case where the ray exits through cell 2):
$$A_6 = (\mathrm dr _x + W\sin \phi - 1)^2 (\cot \phi +  \tan \phi)/2
$$
\end{itemize}

\subsubsection{Cell 8} The area is nonzero if and only if $Q - A < 0$. Let $\beta = (Z - Q)\cot \phi$, $R = Q \tan \phi$, $\eta = \beta - R$.
\begin{itemize}
\item $\tan \phi + Q - Z > 0$:
$$A_8 = \eta^2 \sin \phi \cos \phi /2 
$$
\item $\tan \phi + Q - Z \le 0$: 
$$A_8 = \eta^2 \sin \phi \cos \phi /2 - (\beta - 1) |\tan \phi  + Q - Z|/2
$$
\end{itemize}

\subsubsection{Cell 9} The area is nonzero if and only if $\tan \phi + Q  - Z < 0$.
\begin{itemize}
\item $\tan \phi + Q - A > 0$:
$$A_9 = |\tan \phi + Q - Z| ( (Z - Q) \cot \phi -1)/2
$$
\item $\tan \phi + Q - A \le 0$: Let  $ \iota = |\tan\phi + Q - A| +$\\$ |\tan \phi + Q - Z|$
$$A_9 = \frac
{W \iota \sin \phi - \tan\phi (\tan \phi + Q - A)^2}{2}
$$
\end{itemize}

Now we can retroactively find the the exact area of the intersectional cell.
$$
A_5 = \mathrm dr\times 2W - \sum A_i
$$

\subsection{Generalities}
Not all rays will hit a cell on the left side, nor will it necessarily have an angle between 0 and $\pi/2$. We need to be able to calculate covering fractions for rays entering on any side and with negative $\phi$ values. Luckily, this is simple with translations.

At each place a ray intersects with a cell, the above method outputs a value for each of the cells, so we will represent this as a $3\times3$ matrix.
We can translate this matrix using flips and rotations to get the covering fractions for any possible ray. Rotations adjust for rays entering top, bottom and right sides; flips adjust for negative $\phi$ values ($R(\mathbf{C}_4, \mathbb{Z}_2)$).

\section{Single ray single cell}
We will compare the TegCF prototype to the the current method in the case of the intersection of a single ray with a single cell. Firstly, observe that figure 2 is the covering fraction of the pixel in a 9 cell region. This is a ray with normal $\hat n = [1/\sqrt2,1/\sqrt 2]$ where the ray enters the cell at the bottom corner, $Q = 0$, with $L_{pix} = 1$. Here the covering fraction given by the current method defaults to 1. Observe that the TegCF prototype allocates the radiation more accurately and adds smoothness. 
\begin{figure} [!h]
\hspace*{-.5cm}  
\centering
\includegraphics[totalheight=0.17\textheight]{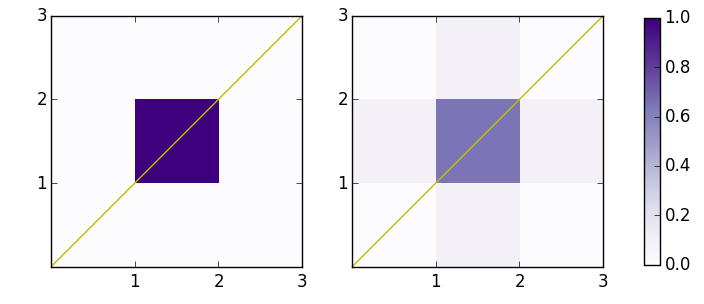}
\caption[short]{Left - current method; right - TegCF: $\hat n = [0.7071,0.7071], Q= 0, L_{pix} =1.0$. Color indicates covering percentage of pixel.}
\end{figure}
\begin{table}[htbp]
\centering
 \begin{tabular}{c c} 
 \multicolumn{2}{c}{Figure 2 Covering Fractions} \\ [0.5ex] 
 \hline
 $f_1$ & 0.0 \\ 
 $f_2$ & 0.0884 \\
 $f_4$ & 0.0884\\
 $f_5$ & 0.6464\\ 
 $f_6$ &0.0884\\ 
$f_8$ & 0.0884\\ 
$f_9$ & 0.0\\ 
$f_{\text{ENZO}}$ & 1.0\\  [1ex] 
 \hline
 \end{tabular}
\end{table}

 Figures 3 and 4 depict two more comparisons of the two methods. Let $f_{\text{ENZO}}$ denote the approximation given by ENZO in the corresponding tables. Both these examples demonstrate how TegCF is lossless with regard to depositing photons, whereas in the current method, there is an inherent loss of photons for any pixel where the area does not overlap over a certain proportion. Also, note the large difference between the approximation, $f_{\text{ENZO}}$, and the exact covering fraction, $f_5$. Cases similar to this cause non-negligable discrepancies in simulations.

\begin{figure} [!h]
\hspace*{-.5cm}  
\centering
\includegraphics[totalheight=0.17\textheight]{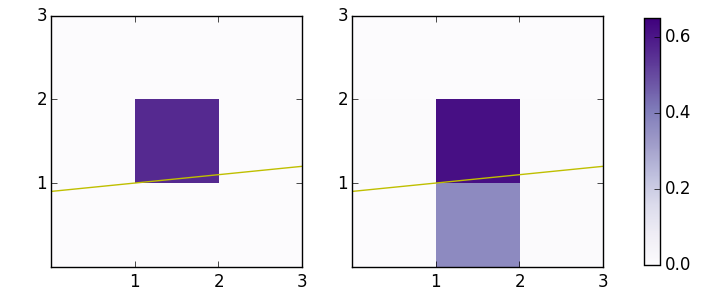}
\caption[short]{Left - current method; right - prototype: $\hat n = [0.9950,0.0995], Q= 0, L_{pix} =0.4$. }
\end{figure}

\begin{table}[!h]
\centering
 \begin{tabular}{c c} 
 \multicolumn{2}{c}{Figure 3 Covering Fractions} \\ [0.5ex] 
 \hline
 $f_1$ & 0.0 \\ 
 $f_2$ & 0.0 \\
 $f_4$ & 0.0050\\
 $f_5$ & 0.6157\\ 
 $f_6$ &0.0050\\ 
$f_8$ & 0.3706\\ 
$f_9$ & 0.0037\\
$f_{\text{ENZO}}$&  0.5625 \\[1ex] 
 \hline
 \end{tabular}
\end{table}
\begin{figure} [!h]
\hspace*{-.5cm}  
\centering
\includegraphics[totalheight=0.17\textheight]{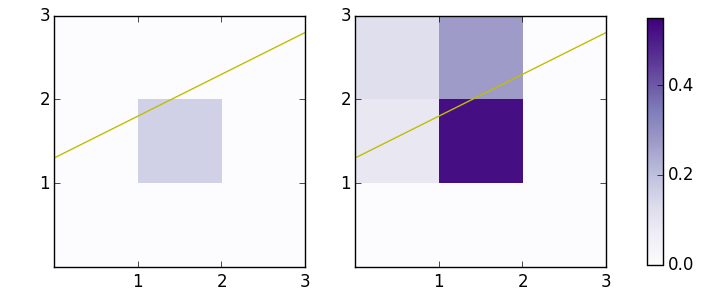}
\caption[short]{Left - current method; right - prototype: $\hat n = [0.8944, 0.4472], Q= 0.8, L_{pix} =1.0$. }
\end{figure}

\begin{table}[!h]
\centering
 \begin{tabular}{c c} 
 \multicolumn{2}{c}{Figure 4 Covering Fractions} \\ [0.5ex] 
 \hline
 $f_1$ & 0.1173 \\ 
 $f_2$ & 0.2708 \\
 $f_4$ & 0.0894\\
 $f_5$ & 0.5224\\ 
 $f_6$ &0.0\\ 
$f_8$ & 0.0\\ 
$f_9$ & 0.0\\
$f_{\text{ENZO}}$ & 0.16\\  [1ex] 
 \hline
 \end{tabular}
\end{table}

\vfill 

\section{Single ray multi-cell}
Here we extend our testing to a single ray spanning a $10 \times 10$ computational grid. We will give two cases for comparison. In figure 5, we examine the extension of the case from figure 2 of the previous section. Again, $L_{pix} = 1$, and $\hat n = [1/\sqrt 2, 1/\sqrt2]$, but it now passes over several cells. Now to further compare the two, we will take the sum of the covering fractions in all cells. We have $F_{\text{ENZO}} = 10$ and $F_{\text{TegCF}}=  9.645$. The discrepancy here results from the computational grid cutting short the cells [11, 10], [10,11], [-1,0], [0, -1]. With these additions, the totals are equal, but the allocation is much smoother in with TegCF.

\begin{figure} [!h]
\hspace*{-.5cm}  
\centering
\includegraphics[totalheight=0.17\textheight]{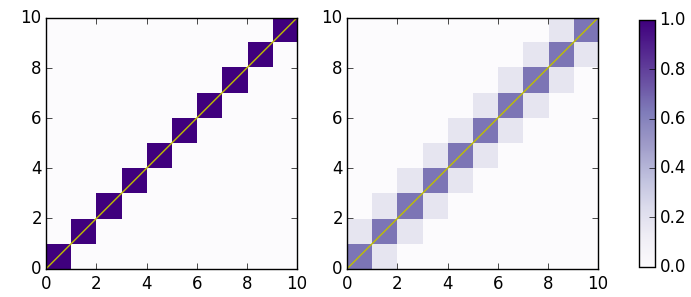}
\caption[short]{Left - current method; right - prototype: $\hat n = [0.7071,0.7071], L_{pix} =1.0$. }
\end{figure}

Figure 6 shows an example wherein TegCF greatly outperforms its counterpart in precision. The ray has total width $L_{pix} = 1$ and a normal of [0.1961, 0.9805]. $F_{\text{ENZO}} = 6.0$, $F_{\text{TegCF}}  = 8.7742$. Thus, there is a loss of a full $2.7742$ pixels which should have been deposited.

\begin{figure} [!h]
\hspace*{-.5cm}  
\centering
\includegraphics[totalheight=0.17\textheight]{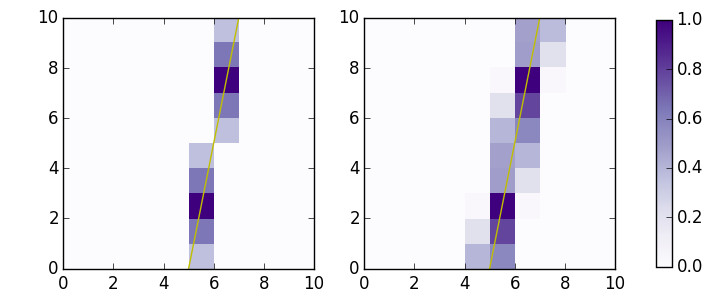}
\caption[short]{Left - current method; right - prototype: $\hat n = [ 0.1961,0.9805], L_{pix} =1.0$. }
\end{figure}

\section{Multi-ray multi-cell}
These are examples of what radiation simulations could look like on a small-scale. We assume we are far from the point source, giving us parallel rays. We again give two cases for comparison. The ``checkerboard" in figure 7 is the extension of figures 2 and 5, with rays spaced $2\Delta c$ apart. This demonstrates how smoothness will counteract artifacts in renderings of radiation. With the correction described in section 4 (but on all sides), the sums of the covering fractions are equal (i.e. lossless in a worst case scenario).

\begin{figure} [!h]
\hspace*{-.5cm}  
\centering
\includegraphics[totalheight=0.17\textheight]{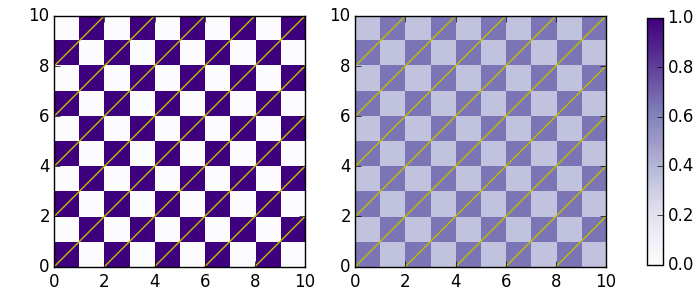}
\caption[short]{Left - current method; right - prototype: $\hat n = [0.1961,0.9805], L_{pix} =1.0$. }
\end{figure}

Our next case is the most drastic difference. It is almost complete coverage using the prototype with almost vertical rays, which is a common occurrence in ENZO. See figure 8 for the illustration. As you can see, even on a very small scale, we have artifacts on the left. In addition to this, there is a massive loss in information. The sum of the covering fractions from the ENZO method is $F_{\text{ENZO}} = 15.11663$, and the sum using the TegCF method is $F_{\text{TegCF}}  = 49.9378$. There is over triple the coverage using the new method. These rays are identical, but the difference between the two methods is extraordinary. 

\begin{figure} [!h]
\hspace*{-.5cm}  
\centering
\includegraphics[totalheight=0.17\textheight]{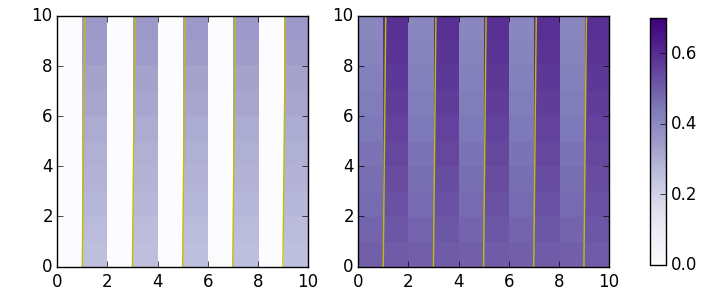}
\caption[short]{Left - current method; right - prototype: $\hat n = [0.1000, 0.0100], L_{pix} =1.0$. }
\end{figure}

\section{Summary}
It should be clear from these examples that the singular case, single ray multi-cell, and multi-ray multi-cell cases can vary greatly with this extension of the geometric correction factor. In this paper we have demonstrated the effectiveness of this extension and its potential value in accurate adaptive ray tracing methods used for radiation simulations. There is room for a large improvement in ENZO without noticeable computation costs. Not only will this counter the artifacts currently displayed, it will add precision and smoothness to general radiative calculations. This prototype was developed in Python, but the key elements should transition over to the C++ source code of ENZO smoothly because of its inclusiveness. 

\section{References}
Abel T. \& Wandelt B. 2001, MNRAS, 330, L53\\
Gorski K. M., Hivon E., Banday A. J., Wandelt B. D., \\ \-\hspace{.5cm} Hansen F. K., Reinecke
M., Bartelmann M., 2005, 
\\ \-\hspace{.5cm}ApJ, 622, 759\\
Wise J. \& Abel T. 2011, MNRAS 414, 3458-3491

\end{document}